\documentclass[10pt,twocolumn,showpacs,prl,preprintnumbers,amsmath,amssymb,aps]{revtex4}
\usepackage{graphicx,dcolumn,bm,color,comment}

\newcommand{\kon}{k_\textrm{\scriptsize{on}}}
\newcommand{\koff}{k_\textrm{\scriptsize{off}}}

\newcommand{\cT}{\mathcal{T}}

\begin{document}
\title{Cooperativity Can Enhance Cellular Signal Detection}
\author{Jianmin Sun}
\email{jmsun@pitt.edu}
\affiliation{Department of Biological Sciences, University of Pittsburgh, Pittsburgh, Pennsylvania 15260, USA}

\author{Michael Grabe}
\email{mdgrabe@pitt.edu}
\affiliation{Department of Biological Sciences, University of Pittsburgh, Pittsburgh, Pennsylvania 15260, USA}

\date{\today}

%
%

\begin{abstract}

Most sensory cells use surface receptors to detect environmental stimuli and initiate downstream signaling. 
Cooperative interactions among sensory receptors is known to play a crucial role in enhancing the sensitivity of biochemical processes such as oxygen sensing by hemoglobin, but whether cooperativity enhances the fidelity with which a system with multiple receptors can accurately and quickly detect a signal is poorly understood. 
We model the kinetics of small clusters of receptors in the presence of ligand, where the receptors act independently or cooperatively. 
We show that the interaction strength and how it is coupled to the dynamics influences the macroscopic observables. 
Contrary to recent reports, our analysis shows that receptor cooperativity can increase the signal-to-noise ratio, but this increase depends on the underlying dynamics of the signaling receptor cluster.

\end{abstract}

\pacs{87.16.Xa, 87.18.Cf, 87.18.Tt, 87.18.Vf}
\maketitle

Biological systems must precisely respond to small changes in environmental cues to carry out essential functions. 
The best studied example of this precision is the cooperative binding of oxygen to hemoglobin, which allows for tight binding of molecular oxygen under high concentrations in the lung and near complete unbinding under low concentrations at distal locations. 
This switch-like response is essential for efficiently carrying out tasks, and cooperativity allows hemoglobin to tune binding and unbinding over a very narrow physiological range. 
Similarly, sensory cells must respond to both small changes in external ligand concentrations and shallow chemical gradients with high accuracy. 
For example, chemotactic bacteria such as \textit{Escherichia coli} can detect and respond to extremely small changes in attractant concentrations near the fundamental Berg-Purcell limit \cite{berg1977physics}, which is set by ligand diffusion. 
It is interesting to imagine that cooperative interactions between receptors may also help increase signal detection, since recent experiments have demonstrated that the bacterial flagellar motor, which is involved in the biochemical pathway for chemotaxis, is highly cooperative \citep{berg, Bai} and cell surface receptors form clusters \citep{Bray}.
Indeed, it has been suggested that clustering allow receptors to interact cooperatively \citep{Mello}, but there is no theoretical basis to suggest that cooperativity helps with signal detection.

\begin{figure}[htbp]
    \includegraphics[width=8cm]{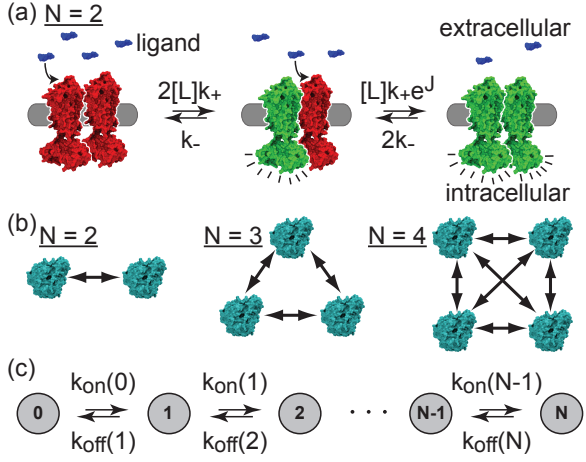}
  \caption{\small (color online) Graphical representations of receptor activation.  
(a) A two receptor model showing one pathway, but not all, for a cluster to move from fully inactive (left) to fully active (right). 
Red receptors are inactive, and they become active (green with dashes) upon ligand binding, whose concentration is [L].
Once one receptor is active, the activation rate for the second receptor is increased by the factor $\alpha = e^J$.
The prefactor 2 in front of the first forward rate and the last reverse rate accounts for the possibility that two receptors can change state.
(b) Schematic diagram of cooperative clusters of size 2, 3 and 4. 
Arrows indicate that two receptors are energetically coupled. 
Unlike nearest neighbor Ising models, all receptors in a cluster interact with all other receptors. 
This is a reasonable assumption for the small cluster numbers considered here, since the proteins likely remain in physical contact.
(c) State diagram with activation and deactivation rates for a cluster composed of $N$ receptors.
Circles represent the number of activated receptors in the cluster, $n$. }
  \label{cartoon}
\end{figure}

Cell signaling is initiated through ligand binding to the extracellular domains of surface receptors, leading to receptor activation (Fig.~\ref{cartoon}(a)). 
The increase in receptor activity is directly proportional to the amplitude of the downstream cellular signal. 
In recent years, there have been several theoretical studies exploring the role of cooperativity in biological signal transduction \cite{bialek2008cooperativity,HuSNR,aquino2011optimal,skoge2011dynamics}. 
Working within the framework of the Monod-Wyman-Changeux (MWC) model, Bialek and Setayeshgar \cite{bialek2008cooperativity} showed that receptor cooperativity lowers the threshold required to sense a change in concentration, bringing the value closer to the Berg-Purcell limit.
Moreover, Hu {\it et al.}~\cite{HuSNR} employed an Ising-type model to show that receptor cooperativity improves gradient sensing in a two dimensional model of the cell embedded in a chemical gradient. 
A crucial concept in signal transduction is the ability to detect the signal over the intrinsic noise in the system, which is known as the signal-to-noise (SNR) ratio. 
This is particularly important in the present case since cooperativity not only sensitizes the receptors' response to a signal, but also amplifies their intrinsic noise due to stochastic fluctuations between activated and inactivated states. 
In this regard, Skoge {\it et al.} \cite{skoge2011dynamics}, also employing a dynamic Ising model, recently showed that cooperativity in chemoreceptor activity slows down receptor activation leading to a reduction in the signal-to-noise ratio compared to non-interacting receptors. 
They showed that this result was true for one and two dimensional Ising models making strong implications that cooperativity does not enhance signal detection. 

The well-known advantages of cooperativity in signal detection juxtaposed with the recent claim that cooperativity fails to enhance the SNR motivated us to more closely examine the role of receptor kinetics on this phenomena. 
In this Letter, we formulate the equilibrium energetics of a small cluster of $N$ interacting receptors using the formalism of statistical mechanics, and the system's dynamics are then studied using classical chemical kinetics, in which forward and reverse rate constants determine the transition rates between distinct structural states of the system (Fig.~\ref{cartoon}). 
The MWC \cite{MWC} and Koshland-Nemethy-Filmer (KNF) \cite{KNF} models make up two seminal schemes for describing the kinetic states involved in the allosteric activation of proteins.
Here, we closely follow the KNF model, which assumes that receptor activation immediately follows from ligand binding, and the activation state of neighboring subunits influence the kinetics of activation and deactivation. 
In general, there are very few constraints placed on a system's dynamics given its energetics and conformational state space, and because of this, the kinetics can be imposed in a number of ways.
For the spin system, it is common to use Glauber dynamics \cite{Glauber} to determine the time evolution of a system.
However, detailed balance only restrains the ratio of the forward to backward rate constants, and many biochemical systems fail to follow either of these dynamics \cite{bohu,schoppa}.
We show here that the relationship between SNR and receptor cooperatively is highly dependent on the manner in which detailed balance is enforced.

We consider small clusters of $N$ interacting receptors, and each receptor can exist in two states: activated or inactivated.
The interaction between receptors is based on the physical proximity of the receptors in a cluster rather than their requirement to fall on a regular lattice, and therefore, the activation/deactivation rate of any given receptor depends on the activation state of all other receptors in the cluster (Fig.~\ref{cartoon}(b)).
The corresponding Hamiltonian is
\begin{equation}
H = -J \sum_{\langle i, j \rangle} \sigma_i \sigma_{j} - h\sum_{i=1}^N
\sigma_i,\label{energy}
\end{equation}
where $J$ is the intrinsic coupling, $\sigma_i = 1$ or $0$ represents the active or inactive receptor states, $\langle i, j \rangle$ denotes links between receptors, and $h$ is related to the energy of ligand binding. 
Specifically, $h = \ln([L]/K_D)$ depends on the bulk ligand concentration $[L]$ and the ligand dissociation constant $K_D$. 
For $J > 0$, the system is cooperative and receptors are biased toward the active state, while $J = 0$ is a model of $N$ independent receptors.

We describe the time evolution of a system of $N$ receptors as shown in Fig.~\ref{cartoon}(c) using the master equation, written here in matrix form
\begin{equation}
\frac{d \mathbf{P}}{dt}=\mathbf{KP},\label{master}
\end{equation}
where $\mathbf{P}$ is a column vector with $N+1$ components $p_n(t)$ and $\mathbf{K}$ is a $(N+1)\times(N+1)$ tridiagonal matrix, whose nonzero elements are given by $K_{nn} = -(\kon(n)+\koff(n))$, $K_{n-1,n} = \kon(n-1)$ and $K_{n, n+1} = \koff(n+1)$. 
Here $p_n(t)$ is the normalized probability of observing $n$ active receptors, $\koff(n)$ is the rate at which a single receptor deactivates given that $n$ receptors are active, and $\kon(n)$ is the rate at which a single receptor is activated given that $n$ receptors are active.
We use reflecting boundary conditions such that $\kon(-1), \koff(0)=0$ and $\kon(N), \koff(N+1)=0$ for the fully inactive and fully activated states, respectively \cite{Gillespie}.
The average number of active receptors at any point in time, $A(t)$, is given by the probability distribution function: $A(t) = \sum_{n=0}^N A_n p_n(t)/N$, where $A_n$ are integer components of the column vector $\mathbf{A} = \{A_n\} = \{0, 1, 2, \cdots ,N\}$.

The ratio of the forward and reverse transition rates between any two adjacent states in Fig.~\ref{cartoon}(c) is related to the energetics of the system through the principle of detailed balance 
\begin{equation}
\frac{\kon(n)}{\koff(n+1)}=\left(\frac{N-n}{n+1}\right)e^{-\Delta E_n}, \label{DB}
\end{equation}
where $\Delta E_n = E_{n+1}-E_{n}$ is the energy difference, in thermal units, between the two states and the values $(N-n)$ and $(n+1)$ account for the number of available receptors for activation or deactivation, respectively.
Our energetic and kinetic formulation in Eqs. \ref{master}-\ref{DB} adheres closely to the approach laid out in Ref. \cite{tchou}. 
According to Eq. \ref{energy}, the energy of the system with $n$ activated receptors is $E_n = -n(n-1)J/2-nh$ leading to $\Delta E_n = -n J-h$.
Given the definition of $h$, the ligand dependent term becomes $\exp(h) = k_+[L]/k_-$, where $k_+[L]$ and $k_-$ are the base activation and deactivation rates for a solitary receptor, respectively.
The dissociation constant $K_D\equiv k_-/k_+$ is defined as the ratio of $k_-$ and $k_+$. Thus, a definition of the rate constants that is consistent with Eq. \ref{DB} is
\begin{eqnarray}
\kon(n)&=&(N-n)k_+[L] e^{\gamma nJ},\nonumber\\
\koff(n+1)&=&(n+1)k_- e^{(\gamma-1) nJ}, \label{rates}
\end{eqnarray}
where $\gamma$ is a phenomenological parameter specifying how the energy of coupling affects the forward and backward transition rates, and we assume that the ligand concentration only influences the activation rate.
When $J = 0$, Eq. \ref{rates} reduces to the standard kinetics for ligand binding to a cluster of non-interacting receptors.
For clusters exhibiting positive cooperativity with $J > 0$, tuning $\gamma$ from 0 to 1 allows us to explore a range of kinetic models. 
For instance, Glauber dynamics results from $\gamma = 1/2$, while a kinetic model in which cooperativity only influences the forward rates and not the reverse rates results from $\gamma = 1$. 

\begin{figure}[htbp]
    \includegraphics[width=8cm]{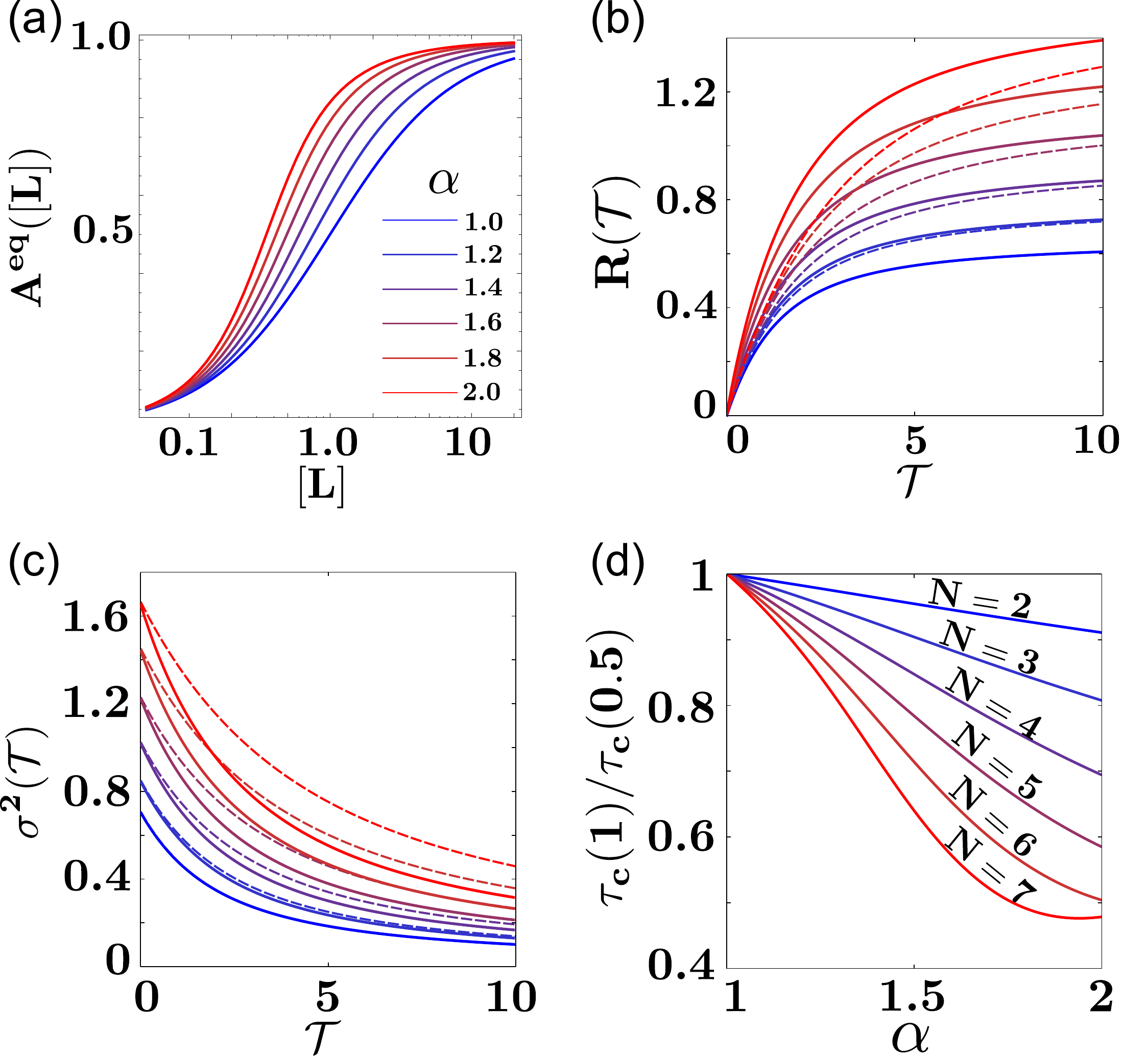}
  \caption{\small (color online) (a) Equilibrium activity of a cluster of $4$ receptors as a function of ligand concentration $[L]$ in the unit of $K_D$, and varying degrees of cooperativity from weak ($\alpha = 1$/blue) to strong ($\alpha = 2$/red). 
(b) Response function for the system of $4$ receptors in panel (a) as a function of observation time ${\cal T}$.
The solid lines correspond to dynamics with $\gamma = 1$, and the dashed lines correspond to $\gamma = 1/2$.  
 For each curve, ${\cal T}$ is scaled by the off rate $1/\koff$. 
 As in panel (a), cooperativity is increased as the color change from blue to red.
(c) Time dependent change in the variance for a $4$-receptor cluster.
All calculations are identical to panel (b).
(d) Ratio of the correlation times for clusters with $\gamma = 1$ dynamics versus $\gamma = 1/2$ dynamics as a function cooperativity and cluster size. }
  \label{Figure2}
\end{figure}

The equilibrium probability distribution, $\mathbf{P}^{eq}=\{p_n^{eq}\}$, can be determined from Eq. \ref{master} by setting the rate of change to zero. 
With this quantity, the equilibrium probability of a single receptor being activated as a function of ligand concentration is $A^{eq}\left([L]\right) = N^{-1}\sum_{n=1}^N A_n p_n^{eq}$.
This can be solved analytically to determine the dependence of receptor activity on ligand concentration and coupling
\begin{equation}
 A^{eq}\left([L]\right) = \frac{1}{N}\frac{\sum_{n=1}^N \dbinom{N}{n} n \prod_{m=1}^n z_m  }{1+\sum_{n=1}^N \dbinom{N}{n} \prod_{m=1}^n z_m } ,
\end{equation}
where $z_m = e^{mJ+h}=([L]/K_D)\alpha^m$ with $\alpha = e^J$.
For four receptors ($N = 4$), $A^{eq}\left([L]\right)$ is shown in Fig.~\ref{Figure2}(a), and we see that as the coupling constant, $\alpha$, increases the activation curve becomes more sigmoidal as expected for highly cooperative transitions such as oxygen binding to hemoglobin. 
Unlike dynamic properties discussed next, the equilibrium properties depend only on the ratio of forward and backward transition rates. 

To determine the kinetic properties of the system after a change in ligand concentration at $t = 0$, we must solve the full master equation in Eq. \ref{master}, which can be done formally as 
\begin{equation}
 \mathbf{P}(t)=e^{\mathbf{K(t-t')}}\mathbf{P}(t'), \label{evolution}
\end{equation}
where the exponentiated rate matrix becomes the conditional probability matrix. 
The average activity of the system, $A(t)$, follows from $\mathbf{P}(t)$, and the change in the system due to the external jump in ligand concentration, $\Delta h=\Delta \ln[L]$, is
\begin{equation}
R({\cal T})= \frac{\Delta A({\cal T})}{\Delta \ln[L]}, 
\end{equation}
where ${\cal T}$ is the observation time after the perturbation.
Assuming that the change in ligand concentration is small, linear response theory \cite{kubo} tells us that the return to equilibrium is proportional to the natural fluctuations in the system, and we can rewrite the time averaged dynamic response function \cite{skoge2011dynamics}, $R$, as
\begin{equation}
R({\cal T})=\frac{1}{\cal T}\int_0^{\cal T} dt \left (\langle A(0)A(0) \rangle -  \langle A(0)A(t) \rangle\right ).
\end{equation} 
The time auto-correlation function can be written as $G(t) = \langle A(0)A(t)\rangle$, which is
\begin{equation}
G(t-t')=\sum_{ij} A_j p(j,t|i,t') \widetilde{A}_i,
\end{equation}
where $p(j,t|i,t')=\left[\exp{\mathbf{K(t-t')}}\right]_{ji}$ is an element of the conditional probability matrix and $\widetilde{A}_i=A_i p_i^{eq}$. 
Using the spectral decomposition, $\mathbf{K}$ can be diagonalized as $\mathbf{K}=\mathbf{Q}^{-1}\mathbf{\Lambda}\mathbf{Q}$, where $\mathbf{Q}$ is formed from the eigenvectors of $\mathbf{K}$ and $\mathbf{\Lambda}$ is a diagonal matrix of its eigenvalues ($\lambda_i$) with
$\lambda_0=0$ and $\lambda_i<0$ for $1\le  i \le N$. 


As shown in Fig.~\ref{Figure2}(b), $R({\cal T})$ increases as cooperativity is introduced into the system.
As expected from a shift in the energetics of the active state, $\alpha = 2$ leads to greater activity than $\alpha = 1$ - the non-interacting case.
Additionally, the response time of the system depends on the kinetic model, with $\gamma = 1$ eliciting faster dynamics than $\gamma = 1/2$, since the forward rate constants are larger. 
The response kinetics of a cluster can be understood more quantitatively in terms of the correlation time $\tau_c$, which characterizes the time required for the response to decay to equilibrium  
\begin{equation}
\tau_c=\frac{1}{\langle\delta A^2\rangle}\int_0^{\infty} dt  \langle\delta A(0)\delta A(t)\rangle\simeq\frac{1}{|\lambda_1|},
\end{equation}
where $\delta A(t)=A(t)-A^{eq}$ and $\lambda_1$ is the largest non-zero eigenvalue of the master equation. 
For a cluster of non-interacting receptors, $\tau_c^{-1} = |\lambda_1| = [L] k_+ + k_-$. 
The addition of cooperativity slows the dynamics, as can be seen qualitatively in Fig.~\ref{Figure2}(b), but as shown in Fig.~\ref{Figure2}(d), $\tau_c$ is much smaller for $\gamma = 1$ dynamics than Galuber dynamics, which have $\gamma = 1/2$.
In fact, this difference becomes more pronounced as the cluster size increases. 
Nonetheless, regardless of the kinetic model, the static response, $R(\infty)$, is independent of $\gamma$, and the solid and dashed lines converge at longer times. 

For a given observation time ${\cal T}$, the signal-to-noise ratio is defined as
\begin{equation}
\textrm{SNR}=\frac{R^2({\cal T})}{\sigma^2({\cal T})},
\end{equation}
where $\sigma^2({\cal T})$ is the average intrinsic noise of the receptor cluster averaged over time ${\cal T}$. 
The response of a multiple independent receptors is always amplified relative to that of the single receptor.
Thus, a cluster of $N$ non-interacting receptors has a signal-to-noise ratio which is $N$ times the SNR of a single receptor.
When cooperativity is introduced into the system, this linear scaling  no longer holds, and the quantity must be calculated using the master equation. 

\begin{figure}[htbp]
    \includegraphics[width=8cm]{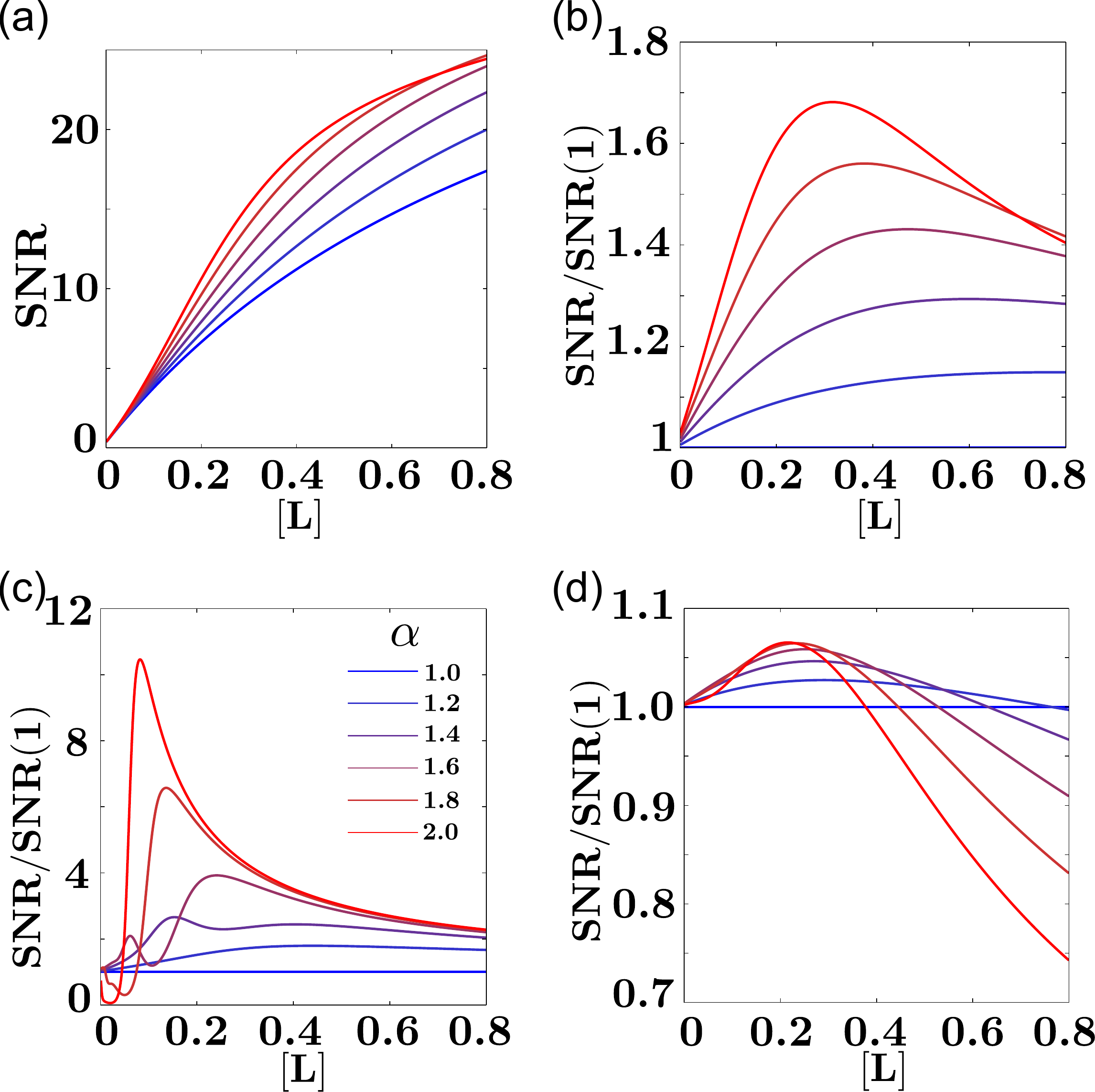}
  \caption{\small (color online) (a) The SNR for a cluster of $4$ receptors as a function of ligand concentration $[L]$ in the unit of $K_D$, with $\gamma=1$ dynamics and varying degrees of cooperativity from weak ($\alpha = 1$/blue) to strong ($\alpha = 2$/red). 
(b) The SNR from panel (a) normalized by the SNR for a cluster of non-interacting receptors. 
(c) Normalized SNR as in panel (b), but for a cluster of $12$ receptors.
(d) Normalized SNR for a cluster of 4 receptors as in panel (b), but with Glauber dynamics ($\gamma=1/2$). }
\label{Figure3}
\end{figure}

The average variance in activity $A$ of a cluster of $N$ receptors, after averaging for time ${\cal T}$, is given in ref. \cite{berg1977physics} as 
\begin{eqnarray}
\sigma^2({\cal T})&=&\frac{1}{{\cal T}^2}\int_0^{\cal T} dt\int_0^{\cal T} dt'  \langle\delta A(t)\delta A(t')\rangle\nonumber\\
&=&\mathbf{A}^\intercal \mathbf{Q}^{-1}{\cal \mathbf{ F({\cal T}) } }\mathbf{Q}\widetilde{\mathbf{A}},
\end{eqnarray}
where $\widetilde{\mathbf{A}}$ is a column-vector with components $\widetilde{A}_i=A_i p_i^{eq}$ and $\mathbf{ F({\cal T}) }$ is defined as
\begin{eqnarray}
\mathbf{ F({\cal T}) } &=&\frac{1}{{\cal T}^2}\int_0^{\cal T} dt\int_0^{\cal T} dt' e^{\mathbf{\Lambda}(t-t')}.
\end{eqnarray} 
$\mathbf{ F({\cal T}) }$ is a diagonal matrix with elements $\mathbf{ F({\cal T}) }_{nn}={\cal F}(\lambda_n{\cal T})$,
where ${\cal F}(x)=2(e^{x}-x-1)/x^2$ is a time dependent function used to describe the dynamics of the variance. 
As the observation time, ${\cal T}$, increases, integration of the signal over multiple transitions from active to inactive leads to a reduction in the variance as shown in Fig.~\ref{Figure2}(c). 
However, the speed with which the variance decays is directly related to the time required for a receptor to cycle between active and inactive states, which is roughly twice the system's correlation time $\tau_c$, as shown previously \cite{kwang, Endres}. 
For large observation times, much greater than $\tau_c$, the variance becomes   
\begin{equation}
  \sigma^2({\cal T})=\frac{2\tau_c}{\cT}\sigma^2(0)\label{Sx0}. 
\end{equation}
In the case of $\alpha=1$, we recover the signal-to-noise ratio of $N$ non-interacting receptors with $R({\cal T})\approx R(\infty)$ for long averaging time,
\begin{equation}
\textrm{SNR}(\alpha=1)=N\frac{{\cal T}k_-}{2} \frac{[L]}{[L]+K_D}.
\end{equation}
We calculated the SNR at time $\cal T$ for clusters containing 4 receptors and no cooperativity, $\alpha=1$ (blue curve in Fig.~\ref{Figure3}(a)). 
As expected, the SNR increases as the ligand concentration is increased.
Next, we introduced increasing amounts of cooperativity by tuning $\alpha$ from $1$ to $2$. 
For these calculations, we assumed model kinetics with $\gamma=1$.
Surprisingly, we realized that cooperativity increased the SNR, as can be most clearly seen in Fig.~\ref{Figure3}(b) where the ratio of the SNR for a cooperative cluster to a non cooperative cluster is always greater than one.
This result indicates that receptors always benefit from cooperativity when $\gamma=1$.
The increased SNR for a cooperative cluster has a strong dependence on the ligand concentration, and its advantage reaaches a maximum around $[L]/K_D=0.3$ for the $\alpha=2$ case.
For a large cluster containing $12$ receptors, the SNR ratio is dramatically increased and for $\alpha=2$ the cooperative cluster is over 10 times more sensitive than the non interacting receptors (Fig.~\ref{Figure3}(c)).
The SNR's dependence on ligand concentration is even more pronounced, and it is shifted to lower concentration values compared to the smaller cluster.
This sharp transition arises from a combination of a more sigmoidal binding curve and a decreased correlation time $\tau_c$.
Therefore, both the equilibrium and kinetic properties play a role in increasing the SNR. 
For small ligand concentrations, cooperativity in large clusters dramatically suppresses the SNR (Fig.~\ref{Figure3}(c)). 
This suppression arises from decreased receptor activation below threshold. 
Finally, in Fig.~\ref{Figure3}(d), we show that a cluster of 4 receptors obeying Glauber dynamics exhibit a reduction in the SNR for high ligand concentrations consistent with a recent report \cite{skoge2011dynamics}. 
Nonetheless, for small ligand concentrations, cooperativity modestly increases the SNR.


Here, we show that cooperativity can increase or decrease the SNR of a receptor cluster depending on the dynamics of receptor activation.
For $\gamma=1$, the SNR can be dramatically increased with respect to a group of non interacting receptors, but this increase is highly dependent on ligand concentration.
Moreover, for $\gamma=1/2$, the SNR is either reduced or only marginally increased with respect to a cluster of non interacting receptors. 
This later result is consistent with previous work \cite{skoge2011dynamics}, but here we have shown that the SNR depends heavily on ligand concentration as well.
Increased sensitivity comes from both the equilibrium properties of the receptors and their intrinsic dynamics.  
The SNR can be increased dynamically by decreasing the correlation time.
Cooperativity reduces the correlation time for clusters with $\gamma=1$ dynamics, while the correlation time increases for clusters obeying Glauber dynamics ($\gamma=1/2$).
Real biological systems have complex dynamics, and it will be interesting to further explore how the detailed kinetics of a particular signalling system influence their macroscopic properties. 

We thank Natalia Jura (UCSF) for insightful conversations. 
This work was supported by NIH Grant No. R21 GM100224-01 (M.G.).

\vfill
\end{document}